\documentclass[twocolumn,prx,showpacs,aps,superscriptaddress]{revtex4-1}

\usepackage[english]{babel}
\usepackage[ansinew]{inputenc}
\usepackage{times}
\usepackage{graphicx}
\usepackage{graphics}
\usepackage{amsmath}
\usepackage{amsfonts}
\usepackage{amssymb}
\usepackage{amsbsy}
\usepackage{dsfont}
\usepackage{epstopdf}
\usepackage{makeidx}
\usepackage{subfigure}
\usepackage{color} 
\usepackage{pgf}
\usepackage{bm}
\usepackage{tikz} \usepackage[normalem]{ulem}
\usepackage{physics}

\begin{document}
\title{Topological Phase Transitions and Quantum Hall Effect in the Graphene Family}

\author{P. Ledwith}
\affiliation{Massachusetts Institute of Technology, Cambridge, Massachusetts 02139, USA}
\affiliation{Center for Nonlinear Studies, MS B258, Los Alamos National Laboratory, Los Alamos, NM 87545, USA}
\affiliation {Theoretical Division, MS B213, Los Alamos National Laboratory, Los Alamos, NM 87545, USA}

\author{W. J. M. Kort-Kamp}
\affiliation{Center for Nonlinear Studies, MS B258, Los Alamos National Laboratory, Los Alamos, NM 87545, USA}
\affiliation {Theoretical Division, MS B213, Los Alamos National Laboratory, Los Alamos, NM 87545, USA}

\author{D. A. R. Dalvit}
\affiliation {Theoretical Division, MS B213, Los Alamos National Laboratory, Los Alamos, NM 87545, USA}

\date{\today}

\begin{abstract}
Monolayer staggered materials of the graphene family present intrinsic spin-orbit coupling and can be driven through several topological phase transitions using external circularly polarized lasers, and static electric or magnetic fields.   We show how topological features arising from photo-induced phase transitions and the quantum Hall effect coexist in these materials, and simultaneously impact their Hall conductivity through their corresponding charge Chern numbers. We also show that the spectral response of the longitudinal conductivity contains signatures about the various phase transition boundaries, that the transverse conductivity encodes information about the topology of the band structure,
and that both present resonant peaks which can be unequivocally associated to one of the four inequivalent Dirac cones present in these materials. This complex optoelectronic response can be probed with straightforward Faraday rotation experiments, allowing the study of the crossroads between quantum Hall physics, spintronics, and valleytronics.

\end{abstract}

\maketitle

\section{Introduction}

The two dimensional staggered semiconductors~\cite{2dsemicon,2dsemicon2,Molle2017} silicene~\cite{SiliceneExperiment}, germanene~\cite{GermaneneExperiment}, stanene~\cite{StaneneExperiment,StaneneExperiment2}, and plumbene~\cite{Yu2017} are monolayer materials made out of Silicon, Germanium, Tin, and  Lead atoms, respectively.  Together with graphene~\cite{GrapheneReview,NunoReview2010}, they make up the group of monolayer honeycomb materials often referred to as the graphene family.  However, as opposed to graphene, these materials have an intrinisic spin-orbit coupling that opens a gap in their electronic band structure. They are also non-planar, with their two inequivalent sublattices lying in two distinct parallel planes, and thus respond to the presence of an out-of-plane static electric field~\cite{EZ1,EZ2,EzawaEZ,NicolEZ}.  Together with a circularly polarized laser, these external fields allow one to tune the gap for each spin and valley, allowing the materials to be driven through several phase transitions~\cite{Photoinduced,MLTopIns,GraFamPT,WKKPT}. Many of the achievable phases possess  topologically nontrivial features that can be characterized by a topological invariant, namely the charge Chern number. On the other hand,  topological states can be also accessed via the quantum Hall (QH) effect  ~\cite{Thouless82}, where a  magnetic field is introduced and the QH Chern number changes depending on the occupation of various Landau levels.  The quantum Hall effect has been studied extensively in graphene~\cite{Goerbig2011} (where it shows an unconventional odd integer quantization originating from the quantum anomaly of the zeroth Landau level in a relativistic spectrum~\cite{UnconventionalHall,GraHallCond}) and in the other members of the graphene family~\cite{MagnetoOptSilPRL,Ezawa2012}. 

Here, we develop a unified and comprehensive study of the interplay between topological features arising from the quantum Hall effect and photo-induced phase transitions in 2D staggered semiconductors. 
Photo-induced and quantum Hall Chern invariants simultaneously manifest themselves in the DC Hall conductivity, resulting in a complex optoelectronic phase diagram possessing a wealth of phase transitions. We discover that doping the monolayer leads to a shift of the phase diagram, allowing to perfectly replicate all photoinduced topological boundaries without the need of circularly polarized light, a phenomenon
that can be traced back to the anomalous nature of the zeroth Landau level. We also demonstrate that the frequency dispersion of the optical conductivity tensor presents several resonances imprinted with signatures of the topologically non-trivial electronic states. Finally, we show that Faraday rotation measurements is a suitable technique to demonstrate the co-existence of Hall effects of distinct origin in the graphene family materials. 


\section{Optical response of the graphene family}

\begin{figure}
\centering
\includegraphics[width=1.0\linewidth]{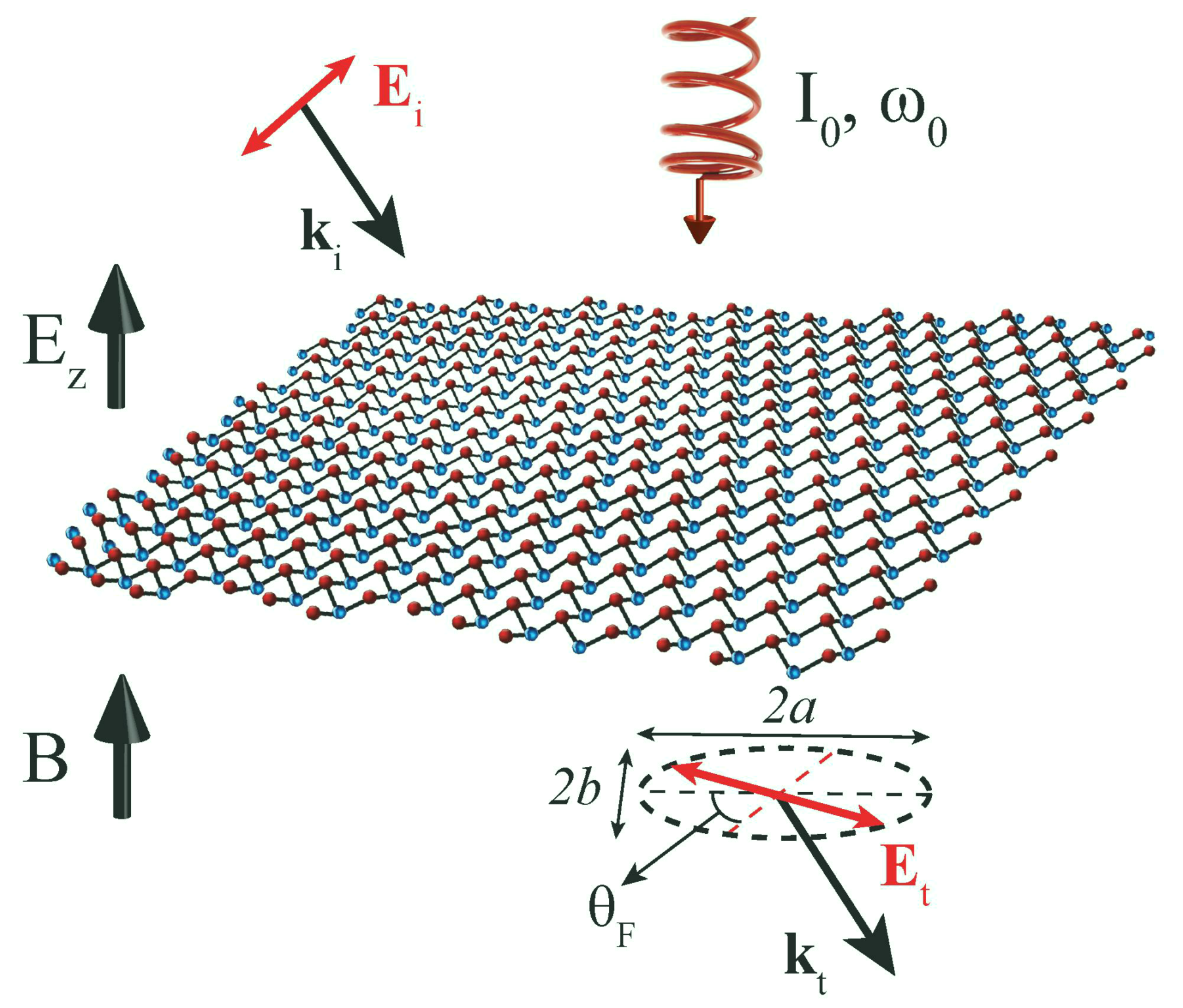}
\caption{Schematics of the system considered: a  staggered monolayer of the graphene family exposed to out-of-plane static electric and magnetic fields together with a normally incident circularly polarized laser.  Optoelectronic properties of the system can be probed with Faraday rotation measurements, where incident linearly polarized light becomes elliptically polarized and undergoes a rotation of the polarization plane after transmission through the monolayer.}
\label{Fig1}
\end{figure}

Let us begin with the Hamiltonian for members of the graphene family, found through the use of a tight-binding model and subsequent low energy expansion, including the effects of a circularly polarized laser and electric field ~\cite{Photoinduced,GraFamPT,MLTopIns},
$\hat{H}^{\eta}_s =v_F(\eta p_x \hat{\tau}_x + p_y\hat{\tau}_y)+\Delta^\eta_s \hat{\tau}_z$, 
where $\Delta^{\eta}_s = -\eta s \lambda_{\text{SO}} + e\ell E_z + \eta \Lambda$ is half the mass gap.
Here, $\hat{\tau}_i$ are Pauli matrices, $\mathbf{p} = (p_x,p_y)$ is the momentum for particles around points $K$ ($\eta = +1$), $K'$($\eta = -1$) and spin $s = \pm 1$, $v_F = \sqrt{3} d t/2\hbar$ is the Fermi velocity, where $d$ is the lattice constant and $t$ is the nearest neighbor coupling.  The Dirac mass $\Delta^\eta_s$ has contributions from the spin-orbit coupling $\lambda_{\text{SO}}$, the out-of-plane electric field $E_z$ and a circularly polarized laser (see Fig. \ref{Fig1}).  The spin-orbit couplings for silicene, germanene, stanene and plumbene are $\lambda_{\text{SO}} \approx 3.9, \, 43, \,100, \, 200 ~\text{meV} $~\cite{Photoinduced,GraFamPT,MLTopIns,Yu2017},  respectively.  Terms originating from Rashba physics are ignored because of their comparitively small effect \cite{Photoinduced,MLTopIns}.  The out-of-plane electric field preferences one sublattice over another due to nonzero lattice buckling $2 \ell$ ($\sim 0.46 ,0.66 , 0.80, 3.00  \, \text{\AA}$, for silicene, germanene, stanene, and plumbene, respectively). The circularly polarized laser of intensity  $I_0$ and frequency $\omega_0$ also modifies the band structure~\cite{Photoinduced,GraFamPT,MLTopIns}. Its interaction with the monolayer can be well described through the coupling constant $\Lambda = \pm 8\pi \alpha v_F^2 I_0/\omega_0^3$ provided  $4\abs{\Lambda}\hbar \omega_0/3t^2 \ll 1$ (where $\alpha$ is the fine structure constant). The full Hamiltonian is block diagonal for each of the possible values of $\eta, s = \pm 1$, hence it sufficies to analyze the spectrum of $H^\eta_s$. The resulting eigenenergies can be cast as $\varepsilon_{\pm} = \pm \sqrt{v_{\text{F}}^2|{\bf p}|^2+(\Delta_{s}^{\eta})^2}$. Since the laser opens a gap in the band structure at  energies equal to $n \hbar \omega_0/2$, where $n$ is an integer,  we shall restrict our discussion to $|\varepsilon_{\pm}| \sim \lambda_{\text{SO}} \ll \hbar \omega_0/2$.

\begin{figure*}{}
\includegraphics[width=0.9\linewidth]{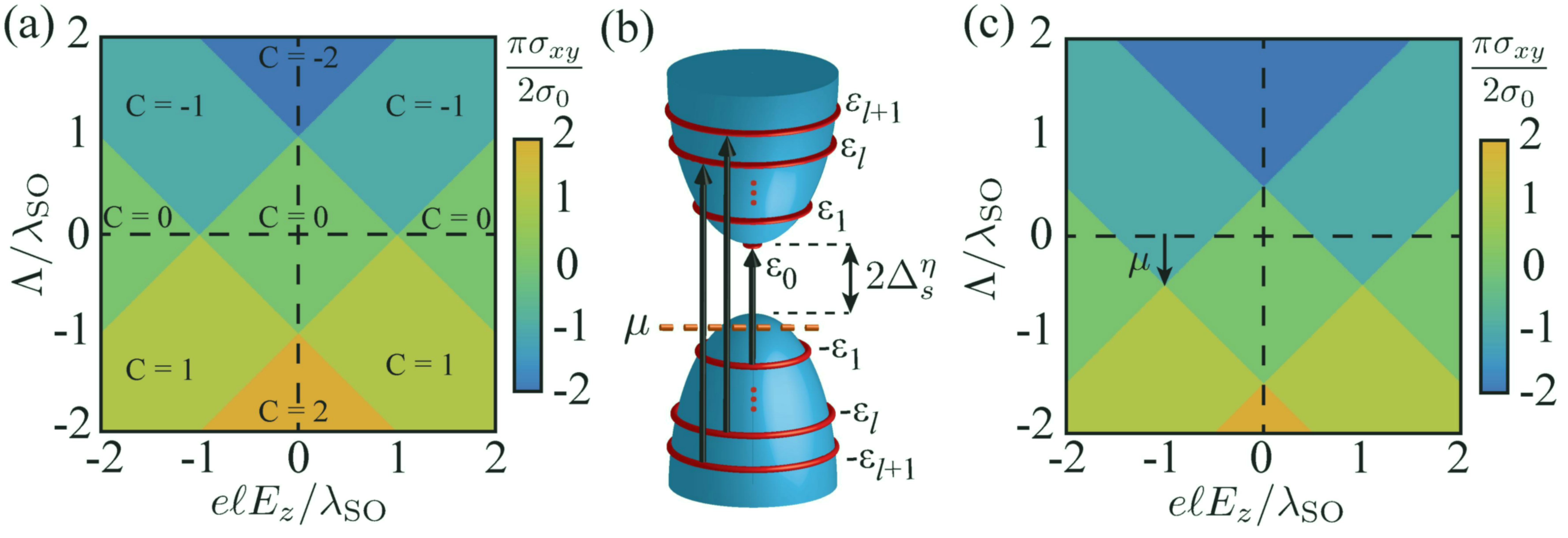}
\caption{
\textbf{a)} Static Hall conductivity $\sigma_{xy}$ plotted in the $(E_z, \Lambda)$ plane at zero magnetic field for a neutral and dissipationless monolayer. The charge Chern number at the phase transition boundaries (lines and points) is given by the average of the charge Chern numbers of the adjacent regions. The phase diagram remains unchanged in the presence of a static magnetic field provided $\mu = 0$ (see text).	\textbf{b)} Relevant transitions for the case $\abs{\mu} < \varepsilon_1$. For illustration purposes only, we consider a cone for which $\varepsilon_0>0$ and choose $\mu < \varepsilon_0$.	
\textbf{c)} Phase diagram for the static $\sigma_{xy}$ for a dissipationless but doped monolayer with 
$\mu / \lambda_{\text{SO}}=0.5$. The magnetic field intensity chosen ($E_B/ \abs{\mu} =100$) is such 
to ensure $\abs{\mu} < \varepsilon_1$ everywhere in the phase plane and for all Dirac cones. Under these conditions, the plot is independent of the particular value of the magnetic field. 
The chemical potential has a role similar to $\Lambda$ and therefore shifts the phase diagram vertically. 
}
\label{Fig2}
\end{figure*}

As the material undergoes a quantum phase transition whenever $\Delta^\eta_s$ vanishes, the tunability of the Dirac mass for each spin and valley allows these materials to exhibit a multitude of different electronic phases, many of which harbor nontrivial topological states \cite{Photoinduced,GraFamPT,MLTopIns}.  The topology is indexed by the charge Chern number 
\begin{equation}
C = - \frac{1}{2} \sum_{\eta,s} \eta \;  \text{sgn}(\Delta^\eta _s).
\label{Ceq1}
\end{equation}
The phase diagram for the graphene family is plotted in Fig. \ref{Fig2}a.  At $E_z = \Lambda = 0$, the material is characterized as a quantum spin Hall insulator with $C=0$.  One can verify that non-zero Chern numbers are generated through the time reversal symmetry breaking. If $E_z$ and $\Lambda$ are both increased, the material remains in the same state as long as $\abs{elE_z} + \abs{\Lambda} < \lambda_{\text{SO}}$.  Along the lines where the previous condition holds as an equality, a single Dirac cone closes giving Chern numbers of $\pm 1/2, \pm 3/2$. At the points where two of these lines intersect, two cones close and the material reaches either the spin valley polarized semimetal ($e\ell E_z/\lambda_{\text{SO}} = 1,\ \Lambda/\lambda_{\text{SO}} =0 $) or  the spin polarized metal ($e \ell E_z/\lambda_{\text{SO}} = 0,\ \Lambda/\lambda_{\text{SO}}=1 $) phase with Chern number $0$ or $\pm1$. In the lowermost and uppermost wedges, the monolayer is an anomalous quantum Hall insulator with $C = \pm 2$, whereas on the leftmost and rightmost wedges, it is a band insulator with $C = 0$. In the regions $\big||\Lambda|-|e\ell E_z|\big|<\lambda_{\text{SO}}<|e\ell E_z|+|\Lambda|$ the material behaves as a polarized spin quantum Hall insulator with $C = \pm 1$.

In order to investigate the quantum Hall effect in the graphene family, let us now assume that a static magnetic field ${\bf B} = B{\bf \hat{z}}$ is applied perpendicularly to the monolayer (see Fig. \ref{Fig1}). The low energy Hamiltonian describing the system is obtained from the above Hamiltonian
through a Peierls substitution $\mathbf{p} \to \mathbf{p} + e\mathbf{A}$, where $\mathbf{A} = -By{\bf \hat{x}}$ is the vector potential in the Landau gauge. Spin splitting due to the Zeeman interaction is ignored owing to its comparatively small effect~\cite{MagnetoOptSilPRL}. Similarly to the case of graphene~\cite{Goerbig2011}, one can solve the energy spectrum and wavefunctions in terms of those of the harmonic oscillator by introducing creation and annihilation operators.  The eigenenergies of the system are
$\varepsilon_n = \text{sgn}(n)\sqrt{(\Delta^\eta_s)^2+\abs{n}E_B^2}$ for $n \neq 0$ ($n$ is an integer), and
$\varepsilon_0 =-\tilde{\eta} \Delta^\eta_s$ for $n=0$,
where $\tilde{\eta} = \text{sgn}(eB)\eta$, and $E_B = \sqrt{2v_F^2 \hbar \abs{eB}}$ is the relativistic analogue of the cyclotron energy. We note that the zeroth level is quantum anomalous: its magnitude is independent of the magnetic field and its sign depends on the particular cone (which means it could be occupied either by electrons or holes). 
We mention that the results we describe in the next sections require $E_B \sim \lambda_{\text{SO}}$, which can be accessed in the graphene family for magnetic fields of the order of a few Teslas. The eigenfunctions associated to the Hamiltonian are
$\ket{n} =  \left(-\tilde{\eta}\,\text{sgn}(n) A^+_n \ket{\abs{n}+\frac{1+\tilde{\eta}}{2}}_{\text{HO}}, 
A^-_n \ket{\abs{n}+\frac{1-\tilde{\eta}}{2}}_{\text{HO}} \right)^T$,
where $\ket{n}_{\text{HO}}$ are the harmonic oscillator eigenstates, $A^\pm_n =  \sqrt{[ \abs{\varepsilon_n}\pm \text{sgn}(n)\Delta^\eta_s ] / 2 \abs{\varepsilon_n}}$ for $n \neq 0$, and $A^\pm_0 = (1 \mp \tilde{\eta})/2$.

The optoelectronic response of the monolayer at  frequency $\omega$ can be characterized by its conductivity $\sigma_{\alpha \beta}$. For the set of parameters we consider in the rest of the paper, we can neglect effects of spatial dispersion and calculate $\sigma_{\alpha \beta}$ using the standard Kubo's approach in the local regime~\cite{WKFaraday,MagnetoOptSilPRB}, resulting in
\begin{equation}
\label{sigmaxxgeneral}
\begin{split}
\frac{\sigma_{xx}}{\sigma_0} = &\frac{iE_B^2}{\pi}  \sum_{\eta,s}\sum_{n,m}  \frac{f_m-f_n}{\varepsilon_n-\varepsilon_m} \\
\times  &\frac{(A^+_mA^-_n)^2\delta_{\abs{n},\abs{m}-\tilde{\eta}} + (A^-_mA^+_n)^2\delta_{\abs{n},\abs{m}+\tilde{\eta}}}{\hbar\omega+\varepsilon_m-\varepsilon_n+i\hbar\Gamma},
\end{split}
\end{equation}
\begin{equation}
\label{sigmaxygeneral}
\begin{split}
\frac{\sigma_{xy}}{\sigma_0} = &-\frac{E_B^2}{\pi}  \sum_{\eta,s}\sum_{n,m}  \eta \frac{f_m-f_n}{\varepsilon_n-\varepsilon_m} \\
\times  & \frac{(A^+_mA^-_n)^2\delta_{\abs{n},\abs{m}-\tilde{\eta}} - (A^-_mA^+_n)^2\delta_{\abs{n},\abs{m}+\tilde{\eta}}}{\hbar\omega+\varepsilon_m-\varepsilon_n+i\hbar\Gamma}\, , 
\end{split}
\end{equation}
and $\sigma_{yy} = \sigma_{xx}$, $\sigma_{yx} = -\sigma_{xy}$. Here, $\sigma_0 = e^2/4\hbar$ is the graphene's universal conductivity,
$f_n  = [e^{(\varepsilon_n - \mu)/k_{\text B} T}+1]^{-1}$ denotes the Fermi Dirac distribution, $T$ is the temperature, $\mu$ is the chemical potential, and $\Gamma$ is the dissipation rate.
 In the next sections we investigate the conductivity tensor for different sets of parameters and show that it contains clear signatures of phase transitions and topology in the graphene family.


\section{Interplay between photoinduced topology and quantum Hall effect}
	
In order to better understand the interplay between the quantum Hall effect and topological phase transitions in the graphene family, we  start our discussion with the DC conductivity of a lossless 2D staggered monolayer. The results we obtain in this regime are an excellent approximation for the conductivity tensor for low frequency and dissipation ($\omega, \Gamma \ll E_B/\hbar$). In this limit, $\sigma_{xx}$ vanishes and $\sigma_{xy}$ is a purely real function. By using that Re$[\sigma_{xy}]$ is symmetric under flipping Landau level indices $n,m$, we can choose $n > m$ and multiply the sum in Eq. (\ref{sigmaxygeneral}) by two. The conductivity can now be interpreted as a sum over transitions $m \to n$ with energy gaps $\varepsilon_n - \varepsilon_m$ and selection rules $\abs{n} = \abs{m} \pm 1$. We also take the limit of zero temperature, allowing us to replace the Fermi-Dirac distribution with Heaviside functions. As we will show in the next sub-section, the static Hall conductivity can be written as 
\begin{equation}
\sigma_{xy} = \frac{2 \sigma_0}{\pi}  \sum_{\eta,s} \theta(\varepsilon_1-|\mu|) \tilde{C}^{\eta,s}_{\rm ph}+
\theta(|\mu| -\varepsilon_1) \tilde{C}^{\eta,s}_{\rm QH},
\label{fulleq}
\end{equation}
 where $\tilde{C}^{\eta,s}_{\rm ph}= (1/2) \text{sgn}(eB)\,\text{sgn}(\varepsilon_0(\Lambda^{\eta}_s)-\mu)$ is the Chern number per cone associated with the photo-induced topology, and $\tilde{C}^{\eta,s}_{\rm QH}= - (1/2) \text{sgn}(eB\mu)(2N^{\eta}_s+1)$ is the Chern number per cone associated with the quantum Hall effect ($N^{\eta}_s$ is the number of filled Landau levels per cone). 

\subsection{Calculation of the DC Hall conductivity}

To evaluate $\sigma_{xy}$, we must determine which transitions are allowed according to the selection rules and the value of $\mu$.  When $\abs{\mu} < \varepsilon_1$, the zeroth Landau level is involved either in the transition $-1 \to 0$ or $0 \to 1$, and there are no intraband transitions.  When, on the other hand, $\abs{\mu} > \varepsilon_1$, the zeroth level no longer contributes, but intraband transitions do.  We therefore split the evaluation of $\sigma_{xy}$ into  two separate cases, $\abs{\mu} < \varepsilon_1$ and $\abs{\mu} > \varepsilon_1$. Note that $\varepsilon_1$ depends on the values of $\eta, s, E_z,$ and $\Lambda$, and so different transitions are possible depending on the cone and location in the phase diagram. 
	
Let us first consider the case  $\abs{\mu} < \varepsilon_1$. When this condition holds,
the allowed transitions in the calculation of the Hall conductivity are $-1 \to 0$ if  $\varepsilon_0 > \mu$  and $0 \to 1$ otherwise, as well as the interband transitions $-l \to l+1$ and $-(l+1) \to l$ for all $\l \geq 1$ [see Fig. \ref{Fig2}b].   All other transitions are either Pauli-blocked or forbidden by the selection rules.  Therefore we can write
$\sigma_{xy}(\Delta^\eta_s) = \sigma_{xy}^{(0)}(\Delta^\eta_s) + \sum_{l \geq 1} \sigma_{xy}^{(l)}(\Delta^\eta_s)$,
where $\sigma_{xy}^{(0)}(\Delta^{\eta}_s)= \sigma_0  E_B^2 \text{sgn}(eB)\,\text{sgn}(\varepsilon_0 - \mu)/ [\varepsilon_1(\varepsilon_1+\text{sgn}(\varepsilon_0 - \mu)\varepsilon_0)]$
is the contribution to the Hall conductivity due to the transition involving the zeroth energy level,  
and 
$\sigma_{xy}^{(l)}(\Delta^{\eta}_s) = \sigma_0E_B^2 \text{sgn}(eB) \, \varepsilon_0 / [\varepsilon_l\varepsilon_{l+1}(\varepsilon_l+\varepsilon_{l+1})]$ is the corresponding contribution due to the $l$ and $l+1$ Landau levels. 
The sum over $l$ can be evaluated analytically by noting that $(\varepsilon_l+\varepsilon_{l+1})^{-1} = (\varepsilon_{l+1}-\varepsilon_{l})/E_B^2$ and concluding that it telescopes, resulting in $\sum_{l \geq 1} \sigma_{xy}^{(l)}(\Delta^\eta_s)  = \text{sgn}(eB)\varepsilon_0/\pi\varepsilon_1$. The final result for the Hall conductivity for a given cone that satisfies the condition $\abs{\mu} < \varepsilon_1$ is 
\begin{equation}
\sigma_{xy}(\Delta^{\eta}_s)= \frac{\sigma_0}{\pi}  \text{sgn}(eB)\,\text{sgn}(\varepsilon_0-\mu)  \equiv 
\frac{2 \sigma_0}{\pi} \tilde{C}^{\eta,s}_{\rm ph},
\end{equation}
where $\tilde{C}^{\eta,s}_{\rm ph}$ is the photo-induced Chern number per cone introduced in Eq.(\ref{fulleq}), which is independent of the magnitude of the magnetic field. A possible way to understand this result is by means of a quantum field theory approach, in which the conductivity can be computed as a sum over filled Landau levels, where negative energy levels are treated as positive ones that can be occupied by holes \cite{UnconventionalHall}.  From this perspective, it is clear that only the zeroth Landau level contributes to the conductivity for $\abs{\mu} < \varepsilon_1$. This is a more physical reason why in the previous approach the sum over transitions telescopes and leaves only the zeroth level contribution.
		 
When the stronger condition $\abs{\mu} < E_B$ is satisfied, then $\abs{\mu} < \varepsilon_1$ for all Dirac cones and everywhere in the phase diagram.  In this case the total conductivity is given by 
\begin{equation}
\sigma_{xy}= \frac{2\sigma_0}{\pi} \sum_{\eta,s} \frac{\text{sgn}(eB)\,\text{sgn}(\varepsilon_0(\Lambda)-\mu)}{2} \equiv \frac{2\sigma_0}{\pi} \tilde{C}_{\rm ph},
\label{generalizedChern}
\end{equation}
where in the last equality we have used that the DC Hall conductivity can always be written in terms of a topological invariant 
\cite{Thouless82} and defined a global photo-induced Chern number $\tilde{C}_{\rm ph}$ for our problem with the magnetic field.
We note that when the monolayer is neutral ($\mu = 0$), using the expression for $\varepsilon_0$
it follows that the photo-induced generalized Chern number $\tilde{C}_{\rm ph}$ is identical to that defined before in the absence of magnetic field Eq.(\ref{Ceq1}), $\tilde{C}_{\rm ph}=C$, and hence we recover the same phase diagram for the Hall conductivity as in the case $B=0$~\cite{Photoinduced,MLTopIns,GraFamPT} (see Fig. \ref{Fig2}a).  This result holds for any magnetic field as long as $\mu = 0$, since $\varepsilon_0$ is independent of the magnitude of $B$ in a relativistic spectrum, as already mentioned.  

For $\mu \neq 0$,  one can verify that $\varepsilon_0(\Lambda) - \mu = \varepsilon_0(\Lambda+\text{sgn}(eB)\mu)$. Therefore, as long as  $\abs{\mu} < E_B$ so that Eq.(\ref{generalizedChern}) holds, 
doping the monolayer shifts all phase boundaries vertically, as depicted in Fig. \ref{Fig2}c. As a result, we can conclude that the chemical potential has a similar effect to that of the  circularly polarized laser, allowing us to replicate all photoinduced phase transition boundaries even for $\Lambda = 0$. 
Thus, in this regime, the chemical potential can serve as an alternative to the high frequency external laser, which is experimentally very difficult to implement since the intensities needed to probe the various topological phase transitions are too large. 
For example, for a green laser impinging on a silicene monolayer
(in this case the condition $\hbar \omega_0 \gg 2 \lambda_{\text{SO}}$ for the validity of the Hamiltonian is well satisfied), one would need a laser intensity as high as  $2.3 \times 10^{14} \, \text{W}/\text{m}^2$ in order to probe regions of the phase diagram with 
$\Lambda= \lambda_{\text{SO}}$.  Such high laser intensities would quickly heat up the sample to  $k_B T > \lambda_{\text{SO}}$, blurring the topological phases altogether (see, e.g., \cite{GraFamPT}). The above results then show that a non-zero chemical potential in addition to a weak magnetic field such that 
$\abs{\mu} < E_B$, (e.g. $\mu=0.5 \lambda_{\text{SO}}$ and $B \sim 0.1 \, \text{T}$ suffices for silicene) would allow experimentalists to probe the different topological phases of these materials ~\cite{Photoinduced,MLTopIns,GraFamPT}, which was until now unrealistic.
\begin{figure}[t]
\includegraphics[width=1\linewidth]{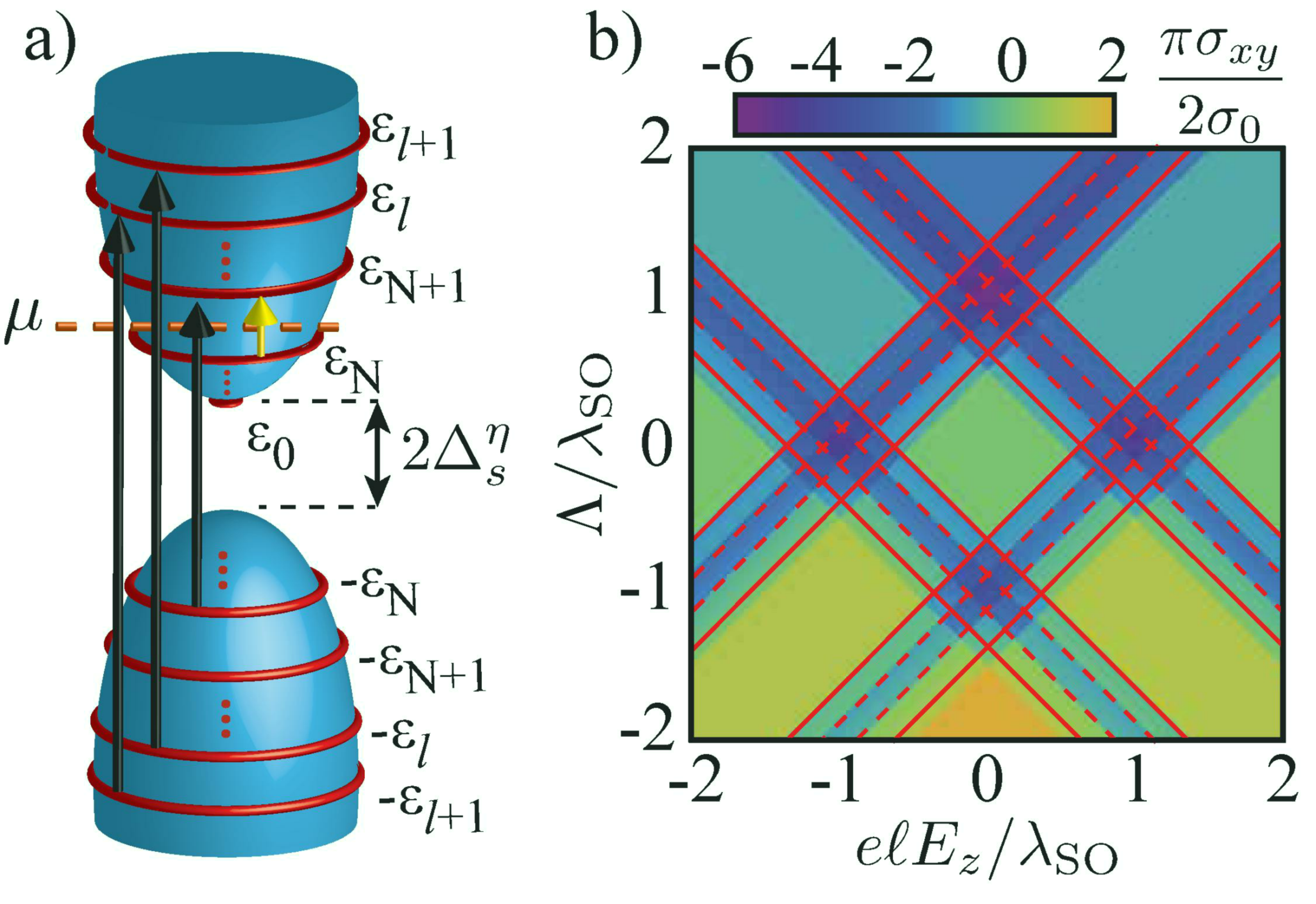}
\label{Fig3}
\caption{\textbf{a)} Relevant transitions for the case $\abs{\mu} > \varepsilon_1$, such that $\varepsilon_N < \mu < \varepsilon_{N+1}$.  The interband transitions are shown with black arrows, and they include the lone transition $-N \to N+1$ as well as the pair of transitions $-l \to l+1$ and $-(l+1) \to l$ for all $l \geq N+1$.  Also shown is the intraband transiton $N \to N+1$, depicted with a gold arrow.
\textbf{b)} Static Hall conductivity $\sigma_{xy}$ plotted in the $(E_z,\Lambda)$ plane for $\mu/\lambda_{\text{SO}}=0.5$, $E_B /\lambda_{\text{SO}}=0.34$, and a dissipationless monolayer.  $N=1$ regions (between adjacent  solid and dashed lines) and $N=2$ regions  (between adjacent dashed lines) appear near the unshifted phase boundaries, where the Dirac mass is sufficiently small.  The shifted boundaries seen in Fig. 2c are still present. 
}
\end{figure}

Let us now consider the situation when $|\mu| > \varepsilon_1$, where the allowed transitions depend on the last filled Landau level per cone
$N^{\eta}_s = \Theta(\mu^2-(\Delta^\eta_s)^2 )\, {\rm Floor}  \left[(\mu^2-(\Delta^\eta_s)^2)/E_B^2\right]$. Note that $N^{\eta}_s > 0$ as we assume $\abs{\mu} > \varepsilon_1$. In order to ease notation, in the following we will drop the indexes $\eta$ and $s$ from $N^{\eta}_s$, but the reader should remember that $N$ depends on cone and on the location in phase space. The allowed transitions for $\mu >0$ are the interband transition $-N \to (N+1)$, the interband transitions $-l \to l+1$ and $-(l+1) \to l$ for all $\l \geq N+1$, and the intraband transition $N \to (N+1)$ [see Fig. 3a].  For $\mu <0$ the sign of the level index and the direction of the transition are both flipped, in accordance with Pauli blocking and selection rules.  
The inter- and intra-band transitions can be computed following similar techniques as above, resulting in the full DC Hall conductivity tensor per cone when $|\mu| > \varepsilon_1$ holds, namely
\begin{equation}
\sigma_{xy}(\Delta^\eta_s) = - \frac{\sigma_0}{\pi} \text{sgn}(eB\mu)(2N+1) \equiv \frac{2 \sigma_0}{\pi} \tilde{C}^{\eta,s}_{\rm QH},
\end{equation}
where $\tilde{C}^{\eta,s}_{\rm QH}$ is the quantum Hall Chern number per cone introduced in Eq.(\ref{fulleq}). Summing over spin and valley, this result reproduces the relativistic Hall effect in gapless graphene ($\Delta^\eta_s \to 0$), for which the Chern number is
$C_{\rm QH}=- \text{sgn}(eB\mu) (4N+2)$ \cite{UnconventionalHall}.

Finally, we combine the results of case $|\mu| < \varepsilon_1$ (for which $N=0$) with those of case 
$|\mu| > \varepsilon_1$
(for which $N \geq 1$), and write a simple expression for the full DC Hall conductivity per cone in the presence of a static electric field, a circularly polarized laser, and a magnetic field, namely
\begin{equation}
\sigma_{xy}(\Delta^\eta_s) = - \frac{\sigma_0}{\pi}\text{sgn}(eB\mu) \left\{ 2N- \text{sgn}(\mu)\text{sgn}[\varepsilon_0(\Delta^{\eta}_s)-\mu] \right \}.
\label{fullDCHall}
\end{equation}
This result is identical to Eq.(\ref{fulleq}) after summing over spin and valley indices.
We also see that this fits our physical picture of electrons (holes) occupying positive energy Landau levels, where the zeroth level has half the degeneracy and an anomalous sign.
		
\subsection{Co-existence of Hall effects in the graphene family}		
		
We now discuss how $\sigma_{xy}$ depends on the location in the $(E_z, \Lambda)$ phase plane at nonzero chemical potential and magnetic field.  When $\mu^2/E_B^2 < 1$, we have that $N=0$ everywhere in the phase plane, and then
$\pi \sigma_{xy}/2 \sigma_0$ is simply equal to $\tilde{C}_{\rm ph}$ with a vertical shift along the $\Lambda$ axis of magnitude $\abs{\mu}$, as discussed before and shown Fig. 2c.  If, however, $\mu^2/E_B^2 > 1$, then sufficiently close to the unshifted boundaries (where the gap goes to zero for a particular cone) $N>0$ contributions arise, and 
$\pi \sigma_{xy}/2 \sigma_0$ is a weighted combination of $\tilde{C}^{\eta,s}_{\rm ph}$ and $\tilde{C}^{\eta,s}_{\rm QH}$ as in Eq.(\ref{fulleq}). This results in a multitude of phase transitions arising from the usual quantum Hall effect ($N\geq 1$) and the photo-induced topological phases ($N=0$), see Fig. 3b. For the chosen parameters in the figure, regions between adjacent dashes lines correspond to situations in which at least one Dirac cone has $N=2$, while those between adjacent parallel solid and dashes lines to situations in which at least one Dirac cone has $N=1$. Squared regions bounded by two solid and two dashed lines correspond to cases in which, out of the four Dirac cones, one has $N=2$, another $N=1$, and the remaining two have $N=0$. Note that the shifted boundaries due to the photo-induced phase transitions also appear in the figure.
Furthermore, in some regions of phase space the Hall conductivity vanishes due to a cancellation between $\tilde{C}^{\eta,s}_{\rm ph}$ of some cones and $\tilde{C}^{\eta,s}_{\rm QH}$ of other ones. 

\begin{figure}[t]
\includegraphics[width=0.9\linewidth]{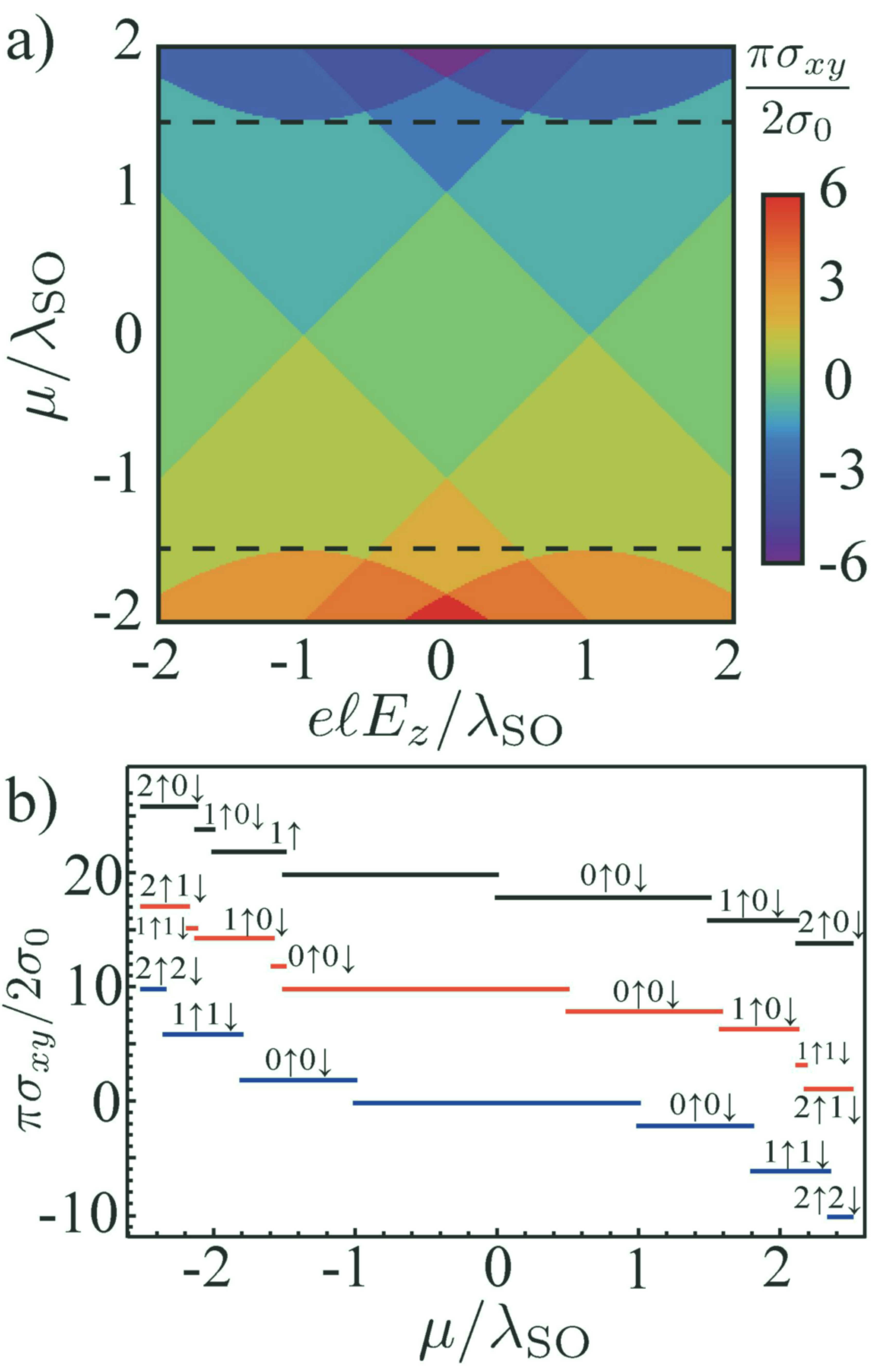}
\caption{{\bf a)} Static Hall conductivity $\sigma_{xy}$ plotted in the $(E_z, \mu)$ plane for $E_B / \lambda_{SO} =1.5$, $\Lambda = 0$, and a dissipationless monolayer. The horizontal strip in between dashed lines corresponds to $\abs{\mu} < E_B$. {\bf b)} Hall conductivity as a function of doping displaying the interplay between photo-induced phase transitions and the quantum Hall effect for $\Lambda/\lambda_{SO}=$ 0 (blue), 1/2 (red), and 1 (black). We choose $E_z=0$, so that Dirac cones are degenerate in the valley index. The pair of labels in each plateau show the last filled Landau level for spin up and down. For cases with $|\mu|<|\varepsilon_0|$ for all four cones, plateaus are not labeled. When this condition is fulfilled for just two cones, the corresponding plateau has a single label for the last filled Landau level with spin up (see the $1\!\!\uparrow$ black plateau on the top-left corner). 
}
\label{Fig4}
\end{figure}

Alternatively, we can plot $\sigma_{xy}$ in the $(E_z,\mu)$ plane as well, fixing $\Lambda = 0$. As depicted in Fig. 4a,
in the horizontal strip defined by $\abs{\mu} < E_B$ (in between the horizontal dashed lines, where $N=0$ for all Dirac cones), the plot is identical to the original plot for $\mu=0$ in the $(E_z,\Lambda)$ plane of Fig. 2a, indicating that doping is a perfect substitute for the laser in this regime. In regions where $|\mu| > \sqrt{E_B^2 + (\eta s \lambda_{SO} - e \ell |E_z|)^2}$, the behavior begins to change due to Dirac cones with $N \neq 0$. Here, four hyperbola-like curves open, defining the boundaries between the photo-induced topological phases and regions where intraband transitions contribute, for particular cones.  For nonzero $\Lambda$, the topological boundaries are shifted vertically and the hyperbolas undergo valley splitting, resulting in eight different hyperbola-like curves (not shown). Fig. 4b shows the Hall conductivity as a function of doping for fixed $E_z=0$ and different values of $\Lambda$. In all cases we observe a ladder-like behavior characteristic of the quantum Hall effect. For $\Lambda/\lambda_{SO}=0$ (which is a vertical cut of Fig. 4a), the three central plateaus correspond to the physics of the case
$|\mu| < \varepsilon_1$, while the outer four plateaus arise from Landau levels with $N>0$. Because in this case the mass gap is degenerate for all four cones, a given plateau has contributions from a unique Landau level. The effect of $\Lambda>0$ is two-fold: i) to shift the central plateaus to the left (due to $\Lambda$ and $\mu$ having the same role for the $N=0$ plateaus), and ii) to enable the interplay between the photo-induced topology and the quantum Hall effect (e.g., plateaus with $N\neq 0$ for spin up and $N=0$ for spin down).


\section{Finite frequency behavior}

\begin{figure*}[t]
\includegraphics[width=1\linewidth]{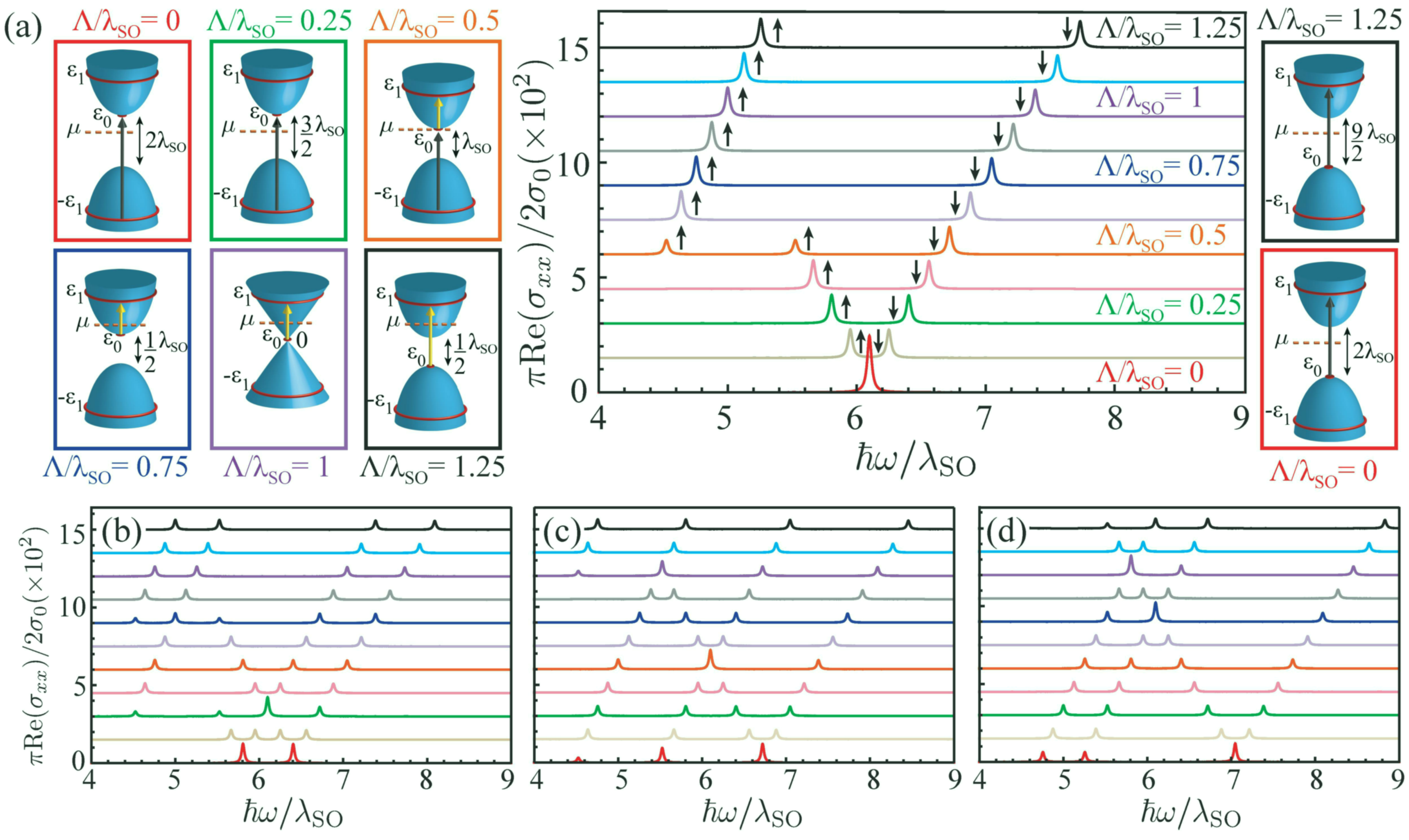}
\caption{{\bf a)} The central panel shows the real part of $\sigma_{xx}(\omega)$ plotted versus frequency for various values of $\Lambda/\lambda_{SO}$ for the case $E_z=0$, which is valley-degenerate. 
The arrows correspond to spin up and down cones. The left subpanels show the allowed transitions for spin up for the corresponding $\Lambda/\lambda_{SO}$ values, while the right subpanels show the allowed transitions for spin down for the selected two extreme values of $\Lambda/\lambda_{SO}$.
{\bf b-d)} Same as the central panel in {\bf a)} but for non-zero electrostatic field, $e \ell E_z/\lambda_{SO}=0.25$, $0.5$, and $0.75$, respectively. Parameters are $\mu/\lambda_{\text{SO}}=0.5$, $E_B/\lambda_{\text{SO}}=5$ (so that $|\mu| < \varepsilon_1$ always), and $\hbar \Gamma/\lambda_{SO}=0.02$.
For clarity, all curves in panels (a-d) are vertically shifted by $1.5$ with respect to each other.
}
\label{Fig5}
\end{figure*}

We now turn our attention to the case of finite frequency in order to show that, also in this case, the conductivity tensor displays signatures of topological phase transitions. For simplicity, we will restrict ourselves to frequencies for which only transitions between the 0 and $\pm 1$ Landau levels are involved.
This means that we will be always in the case $|\mu| < \varepsilon_1$.
To this end, we first numerically compute the conductivity tensor at finite frequency and dissipation as given in Eqs. (\ref{sigmaxxgeneral},\ref{sigmaxygeneral}), and show plots of $\text{Re} [\sigma_{xx}(\omega)]$ and
$\text{Re} [\sigma_{xy}(\omega)]$ for various points in the electronic phase space. Results for $\text{Im} [\sigma_{xx,xy}(\omega)]$ can be obtained using the Kramers-Kronig relations. 

In Fig. \ref{Fig5}a we show the impact of topological phase transitions in the frequency dispersion of the real  part of the longitudinal component of the conductivity tensor for $E_z = 0$. Resonances occur when $\hbar \omega$ matches the gap between two Landau levels, the smaller of which is occupied and the larger unoccupied.  
Since $E_z = 0$, resonances are valley-degenerate, and hence we only need to distinguish between cones with up and down spin. 
At $\Lambda/\lambda_{SO} = 0$,  for spin up the allowed transition is $-1 \rightarrow 0$ (see red subpanel on the left side of the figure), while for spin down it is $0 \rightarrow 1$ (see red subpanel on the right side). Since both up and down spin cones have the same transition gap $\varepsilon_1+|\varepsilon_0|$, they have the same resonance (see red curve in the central panel). 
As $\Lambda/\lambda_{SO}$ increases, the transition gap $\varepsilon_1+|\varepsilon_0|$ grows for spin down (see black subpanel on the right side of the figure) since
$|\Delta^{\eta}_{-1}|$ becomes larger and, as a consequence, the spin down resonances move to higher frequencies, as shown on the central panel in Fig. \ref{Fig5}a.
In contrast, for spin up cones, both $\varepsilon_0$ and the gap $\varepsilon_1+ |\varepsilon_0|$ corresponding to the transition $-1 \rightarrow 0$ decrease as $\Lambda$ grows, causing the spin up resonances to move to smaller frequencies (see green subpanel on the right, $\Lambda/\lambda_{SO}=0.25$). 
This continues until the phase transition boundary is reached ($\varepsilon_0= \mu$), at which point the 
$0 \rightarrow 1$  transition also becomes allowed (see orange subpanel, $\Lambda/\lambda_{SO}=0.5$), with a gap equal to $\varepsilon_1 - |\varepsilon_0|$. We therefore expect that, at the phase transition point, the spin up resonance splits into two new resonances separated in frequency by $2 |\mu|/\hbar$ with half the spectral weight of the original resonance (see orange curve in central panel). 
Once $\varepsilon_0$ goes below $\mu$, only the $0 \rightarrow 1$ transition contributes. While $\varepsilon_0>0$, the transition gap is $\varepsilon_1- |\varepsilon_0|$ which grows as 
$\Lambda$ increases (see blue subpanel for $\Lambda/\lambda_{SO} =0.75$ and dark blue curve in central panel). When $\Lambda/\lambda_{SO} =1$, the cones touch, $\varepsilon_0=0$ and the transition gap is equal to $\varepsilon_1$ (see purple subpanel). Further increasing $\Lambda/\lambda_{SO}$, $\varepsilon_0$ hops from the top cone to the bottom cone as it changes sign. This, however, does not correspond to a phase transition, and the transition gap $\varepsilon_1 + |\varepsilon_0|$ just continues to grow with the spin up resonance shifting to higher frequencies. 
As shown in Figs. \ref{Fig5}b-d, the situation is more complex when $E_z \neq 0$, since then all cones are in general non-degenerate. However, the same general principles of the resonance hopping by $2|\mu|/\hbar$ and changing direction across a phase transition still apply. In summary, the finite frequency behavior of $\text{Re}[ \sigma_{xx}(\omega)]$ allows to detect the phase transition boundaries in the graphene family electronic phase diagram. 

\begin{figure}[t]
\includegraphics[width=1\linewidth]{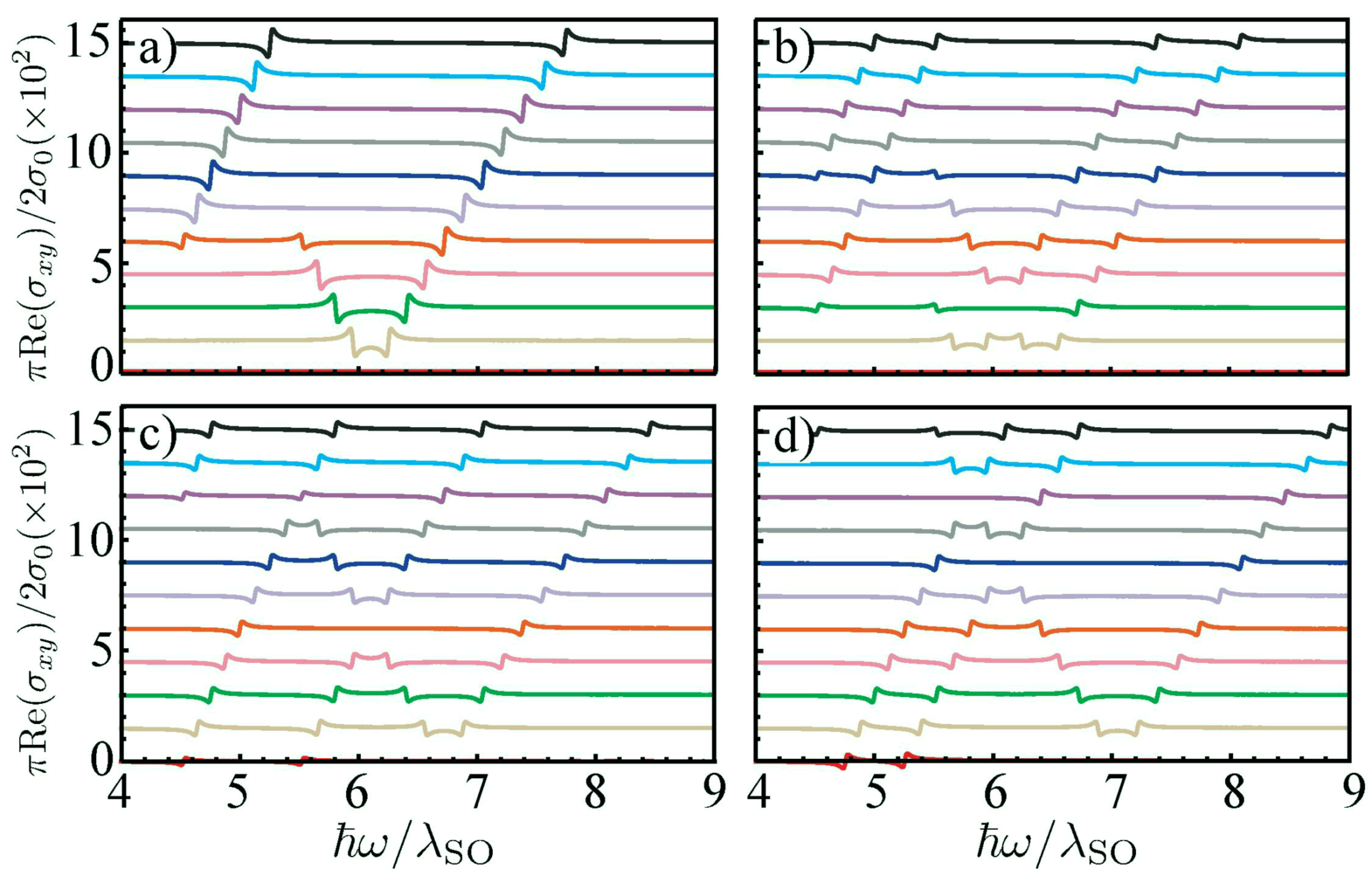}
\caption{Real part of $\sigma_{xy}(\omega)$ plotted versus frequency for various values of $\Lambda/\lambda_{SO}$ for $e \ell E_z/\lambda_{SO}$ equal to {\bf a)} 0,  {\bf b)} 0.25, {\bf c)} 0.5, and
{\bf d)} 0.75. Color scheme, vertically shifting of curves, and parameters are the same as in Fig. 5.}
\label{Fig6}
\end{figure}

Fig. \ref{Fig6} shows $\text{Re}[ \sigma_{xy}(\omega)]$ vs frequency for various values of $\Lambda$ and $E_z$. Apart from featuring resonant and anti-resonant behavior, the overall structure is qualitatively similar to that of $\text{Re}[ \sigma_{xx}(\omega)]$. 
Just as the DC Hall conductivity allows to probe photo-induced topological features of the graphene family materials (see Section III), so it does at finite frequencies.
Indeed, the photo-induced charge Chern number $\tilde{C}_{\rm ph}$ at any $(E_z, \Lambda)$ point can be computed from Fig. \ref{Fig6} by summing the signs of the slopes between adjacent resonances and anti-resonances, accounting appropriately for degeneracy, and multiplying the result by -1/2. As an example, we
consider the case $E_z=0$ (Fig.\ref{Fig6}a). For $\Lambda/\lambda_{SO}=0.25$ (green curve), the two pairs of resonance-antiresonance have opposite slopes, resulting in $\tilde{C}_{\rm ph}=0$. 
For $\Lambda/\lambda_{SO}=0.5$ (orange curve), the two split spin up resonance-antiresonance pairs cancel the contribution of each other, while the spin down resonance-antiresonance has positive slope and degeneracy equal to two, resulting in $\tilde{C}_{\rm ph}=-1$. 
For $\Lambda/\lambda_{SO}=0.75$ (dark blue curve), both resonance-antiresonance pairs have the same slope with degeneracy equal to two, resulting in $\tilde{C}_{\rm ph}=-2$. Analogous analysis can be done for cases with $E_z \neq 0$.  

\section{Discussion}

All the phenomena described above associated with the Hall conductivity $\sigma_{xy}$ can be probed experimentally through Faraday rotation \cite{WKFaraday}. As depicted in Fig. 1, incoming s-polarized light transmitted through the monolayer in general becomes elliptically polarized ~\cite{FaradayRot}. The Faraday rotation angle $\theta_F$ and the minor-to-major axis ratio $b/a$ are given by $\theta_F = \arg(T_+/T_-)/2$ and $b/a = \abs{\chi}$, where 
$\chi = (\abs{T_+}-\abs{T_-})/(\abs{T_+}+\abs{T_-})$ and $T_{\pm} = t_{ss} \pm i t_{ps}$.
Here, $t_{ss}$ and $t_{ps}$ are  the co- and cross-polarized Fresnel transmission coefficients for incoming s-polarized light (see, for example, \cite{ReflTransSpontE} for their expressions). To linear order in the fine structure constant $\alpha$, one obtains $\theta_F \approx  - (Z_0/2) {\rm Re} [ \sigma_{xy}]$ and
$\chi \approx - (Z_0/2)  {\rm Im} [ \sigma_{xy}]$, where $Z_0$ is the vacuum impedance. As we discussed in Section III, for low frequency and dissipation ($\omega, \Gamma \ll E_B/\hbar$) the Hall conductivity is real, so $\chi \approx 0$ (linearly polarized transmission), and $\theta_F$ contains information about the photo-induced and quantum Hall topological invariants (see Eq.(\ref{fulleq})). For the particular case of $|\mu| < E_B$, $\theta_F$ is directly proportional to the photo-induced Chern number per Eq.(\ref{generalizedChern}). 
Using the parameters of Figs 2-3, for a photo-induced Chern number of $\tilde{C}_{\rm ph} = 0, -1, -2$, the Faraday rotation angles are $\theta_F \approx 0, 0.0073$, and $0.0146$ rad, respectively.
For finite frequencies and dissipation, $\theta_F$ will experience all the same resonance/anti-resonance behavior of ${\rm Re} [\sigma_{xy}(\omega)]$ shown in Fig. \ref{Fig6}, and in particular it can also probe the topological features of the monolayer. For example, for $E_z=0$, $\Lambda/\lambda_{SO}=0.25$, $\hbar \Gamma/\lambda_{SO}=0.02$, $E_B/\lambda_{SO}=5$, and for frequencies around
$\hbar \omega / \lambda_{SO} \approx 5.8$, we get values for the Faraday rotation angle as large as $\theta_F \approx 0.36$ rad. Regarding the state of polarization of the transmitted field at finite frequencies, one finds $\chi \approx 0$ for any frequency except near resonances. At $\hbar \omega / \lambda_{SO} \approx 5.8$ we get $|\chi| \approx 0.4$ 
(elliptically polarized light) for the same values of $E_z$, $\Lambda$, $\Gamma$, and $E_B$ as before. The above range of values for $\theta_F$ and $\chi$ should be within experimental reach.

In summary, we have discussed the interplay between photo-induced topological phase transitions and the quantum Hall effect in the graphene family materials. We showed that, in the absence of the external circularly polarized laser, doping these 2D semiconductors below their first Landau level results in an equivalent low-frequency opto-electronic response as for the case with the laser and no magnetic field, thus providing a practical alternative way to probe unusual Hall physics from photo-induced topological phase transitions in the graphene family. Higher values of doping result in a more complex optical response, where such phase transitions co-exist with topological features arising from the quantum Hall effect. We envision that the effects predicted in this work will greatly impact ongoing research in spintronics and valleytronics in emergent van der Waals materials.  


\section*{Acknowledgements}
We are grateful to P. Rodriguez-Lopez and L. Woods for discussions. We acknowledge financial support from the Los Alamos National Laboratory (LANL) Laboratory Directed Research and Development (LDRD) program and the Center for Nonlinear Studies (CNLS).



\begin{thebibliography}{99}

\bibitem{2dsemicon}
A. Castellanos-Gomez, Nat. Phot. {\bf 10}, 202 (2016).

\bibitem{2dsemicon2}
A. J. Mannix, B. Kiraly, M. C. Hersam, and N. P. Guisinger,
Nat. Rev. Chem. {\bf 1}, 0014 (2017).

\bibitem{Molle2017}
A. Molle, J. Goldberger, M. Houssa, Y. Xu, S.-C Zhang, and D. Aknwande,
Nat. Mat. {\bf 16}, 163 (2017).

\bibitem{SiliceneExperiment}
P. Vogt, P. de Padova, C. Quaresima, J. Avila, E, Frantzeskakis, M. C. Asensio, A. Resta, 
B. Ealet, and G. Le Lay, Phys. Rev. Lett. {\bf 108}, 155501 (2012).

\bibitem{GermaneneExperiment}
M. E. D{\'a}vila, L. Xian, S. Cahangirov, A. Rubio, and G. Le Lay,
New Journal of Physics {\bf 16}, 095002 (2014).

\bibitem{StaneneExperiment}
F.-F. Zhu, W.-J. Chen, Y. Xu, C.-L. Gao, D.-D. Guan, C.-H. Liu, D. Qian, S.-C. Zhang, 
and J. F. Jia, Nat. Mat. {\bf 14}, 1020 (2015).

\bibitem{StaneneExperiment2}
S. Saxena, R. P. Chaudhary, and S. Shukla, 
Scientific Reports {\bf 6}, 31073 (2016).

\bibitem{Yu2017}
X.-L. Yu, L. Huang, and J. Wu, 
Phys. Rev. B {\bf 95}, 125113 (2017).

\bibitem{GrapheneReview}
A. H. Castro Neto, F. Guinea, N. M. R. Peres, K. S. Novoselov, and A. K. Geim, 
Rev. Mod. Phys. {\bf 81} 109 (2009).

\bibitem{NunoReview2010}
N. M. R. Peres, 
Rev. Mod. Phys. {\bf 82} 2673 (2010).

\bibitem{EZ1}
N. D. Drummond, V. Z\'olyomi, and V. I. Fal'ko, 
Phys. Rev. B {\bf 85}, 075423 (2012).

\bibitem{EZ2}
Z. Ni, Q.  Liu, K. Tang, J. Zheng, J. Zhou, R. Qin, Z. Gao, D. Yu, and J. Lu, Nanoletters {\bf 12}, 113 (2012).

\bibitem{EzawaEZ}
M. Ezawa, New Journal of Physics {\bf 14}, 033003 (2012).

\bibitem{NicolEZ}
L. Stille, C. J. Tabert, and E. J.  Nicol, Phys. Rev. B {\bf 86}, 195405 (2012).

\bibitem{Photoinduced}
M. Ezawa, 
Phys. Rev. Lett. {\bf 110}, 026603 (2013). 

\bibitem{MLTopIns}
M. Ezawa, Journal of the Physical Society of Japan {\bf 84}, 121003 (2015).

\bibitem{GraFamPT}
P. Rodriguez-L\'opez, W. J. M. Kort-Kamp, D. A. R. Dalvit, and L. M. Woods,
Nat. Comm. {\bf 8}, 14699 (2017).

\bibitem{WKKPT}
W. J. M. Kort-Kamp,
Phys. Rev. Lett. {\bf 119}, 147401 (2017).

\bibitem{Thouless82}
D. J. Thouless, M. Kohmoto, M. P. Nightingale, and M. den Nijs,
Phys. Rev. Lett. {\bf 49}, 405 (1982).

\bibitem{Goerbig2011}
M. O. Goerbig, Rev. Mod. Phys. {\bf 83}, 1193 (2011).

\bibitem{UnconventionalHall}
V. P. Gusynin and S. G. Sharapov, Phys. Rev. Lett. {\bf 95}, 146801 (2005).

\bibitem{GraHallCond}
V. P. Gusynin, S. G. Sharapov, and J. P. Carbotte, Journal of Physics: Condensed Matter {\bf 19}, 026222 (2007).


\bibitem{MagnetoOptSilPRL}
C. J. Tabert and E. J. Nicol, Phys. Rev. Lett. {\bf 110}, 197402 (2013).

\bibitem{Ezawa2012}
M. Ezawa, Journal of the Physical Society of Japan {\bf 81}, 064705 (2012).

\bibitem{WKFaraday}
W.-K. Tse, A. H.  MacDonald, Phys. Rev. Lett. {\bf 105}, 057401 (2010).

\bibitem{MagnetoOptSilPRB}
C. J. Tabert and E. J. Nicol, Phys. Rev. B. {\bf 88}, 085434 (2013).

\bibitem{FaradayRot}
I. V. Fialkovsky and D. V. Vassilevich, Journal of Physics A: Mathematical and Theoretical {\bf 42}, 442001 (2009).

\bibitem{ReflTransSpontE}
W. J. M. Kort-Kamp, B. Amorim, G. Bastos, F. A. Pinheiro, F. S. S. Rosa, N. M. R. Peres, and C. Farina, Phys. Rev. B {\bf 92}, 205415 (2015).

\end{thebibliography}
\end{document}